\documentstyle[twocolumn,floats,aps,prb,epsf]{revtex}
\draft

\title{\boldmath Disorder-driven non-Fermi-liquid behavior in CeRhRuSi$_2$}

\author{Chia-Ying Liu, D.~E. MacLaughlin, and A.~H. Castro Neto}
\address{Department of Physics, University of California, Riverside, California
92521}
\author{H.~G. Lukefahr}
\address{Whittier College, Whittier, California 90608}
\author{J.~D. Thompson, J.~L. Sarrao, and Z. Fisk}
\address{Los Alamos National Laboratory, Los Alamos, New Mexico 87545}

\author{\small(Submitted June 3, 1999; revised version submitted September 14, 
1999)}
\address{\parbox{14cm}{\bigskip\rm\small
We report measurements of the bulk magnetic susceptibility and $^{29}$Si 
nuclear magnetic resonance (NMR) linewidth in the heavy-fermion 
alloy~CeRhRuSi$_2$. The linewidth increases rapidly with decreasing temperature 
and reaches large values at low temperatures, which strongly suggests the wide 
distributions of local susceptibilities~$\chi_j$ obtained in disorder-driven 
theories of non-Fermi-liquid (NFL) behavior. The NMR linewidths agree well with 
distribution functions~$P(\chi)$ which fit bulk susceptibility and specific 
heat data. The apparent return to Fermi-liquid behavior observed below 1 K is 
manifested in the vanishing of $P(\chi)$ as $\chi \to \infty$, suggesting the 
absence of strong magnetic response at low energies. Our results indicate the 
need for an extension of some current theories of disorder-driven NFL behavior 
in order to incorporate this low-temperature crossover.
\\[6pt] PACS numbers: 71.27.+a, 75.30.Mb, 76.60.Cq.}}
\begin{document} \maketitle					

\thispagestyle{myheadings}
\markright{{\small Submitted to {\em Physical Review B}}\hfill {\small
LA-UR-98-4748\quad p.}\hspace{1mm}}

\section{Introduction} \label{sect:intro}

A breakdown of the standard Landau Fermi-liquid theory is signaled in certain 
heavy-fermion metals by anomalies in thermodynamic, transport, and optical 
properties at low temperatures and frequencies.\cite{ITP96} Although exceptions 
exist, the anomalous properties are usually as follows: the Sommerfeld specific 
heat coefficient~$\gamma(T) = C(T)/T$ diverges as $-\ln T$; the magnetic 
susceptibility~$\chi(T)$ varies as $1 - aT^{1/2}$ or diverges as $-\ln T$ or a 
weak inverse power of temperature; the electrical resistivity departs linearly 
with temperature from its $T = 0$ value; and optical conductivity experiments 
in the non-Fermi-liquid (NFL) alloy~UCu$_{3.5}$Pd$_{1.5}$ indicate a transport 
relaxation rate which varies linearly with frequency at low 
temperatures.\cite{DeOt96}

Attempts to understand this NFL behavior invoke one or more characteristics 
common to most such systems, viz., the possibility of an unconventional Kondo 
effect, \cite{Cox87,AfLu91} proximity to a quantum critical point 
(QCP),\cite{Cont93,Mill93,TsRe93} structural disorder,\cite{MDK96} or a 
combination of the latter two.\cite{CNCJ98} Recent experimental work has 
stressed the role that disorder can play. In particular, the observed 
inhomogeneous broadening of copper nuclear magnetic resonance (NMR) lines in 
the NFL alloys~UCu$_{5-x}$Pd$_x$, $x=1.0$ and 1.5, could be described by a 
disorder-induced spatial distribution of local susceptibilities~$\chi_j$. 
\cite{BMLA95,MBL96} Such a susceptibility distribution originates in the 
interplay between structural disorder and many-body effects intrinsic to 
$f$-electron systems, such as the Kondo effect and the 
Ruderman-Kittel-Kasuya-Yosida (RKKY) interaction between the magnetic moments.

It is clear that if an interaction between moments is present the term 
``non-Fermi liquid'' must be used with care, since strictly speaking 
Fermi-liquid theory deals only with the lowest-lying excitations of a system of 
interacting fermions and hence is correct only in the zero-temperature limit in 
the absence of a phase transition. Any kind of magnetic phase transition or 
glassy spin freezing at nonzero temperature invalidates this condition, and NFL 
behavior is no longer a surprise. Following convention in this field, we 
nevertheless continue to designate as NFL systems materials in which the 
properties mentioned above are found at intermediate temperatures, with the 
proviso ``nearly NFL'' applied if there is evidence for a crossover to a new 
state at low temperatures.

In this paper we report measurements of the magnetic susceptibility and 
$^{29}$Si NMR linewidth in the nearly-NFL \cite{GTHM97} heavy-fermion 
alloy~CeRhRuSi$_2$, and consider the data in the light of two such 
disorder-driven scenarios: (1)~the so-called ``Kondo disorder'' picture of 
Bernal {\em et al.\/} \cite{BMLA95} and Miranda, Dobrosavljevi{\'c}, and 
Kotliar, \cite{MDK96} in which the RKKY interaction is disregarded and the 
local susceptibility distribution is associated with a corresponding 
distribution of single-ion Kondo temperatures~$T_K$, and (2)~a recent model by 
Castro Neto and co-workers \cite{CNCJ98} based on the existence of quantum 
Griffiths singularities \cite{Grif69} in a disordered system with RKKY 
couplings which is close to a QCP. \cite{ITP96,Mill93} In the latter case 
Kondo and RKKY phenomena compete with each other in the random environment, 
and the susceptibility is associated with fluctuations of magnetic 
clusters.\cite{CNCJ98} Both of these models have been shown to account for the 
observed susceptibility and NMR broadening in UCu$_{5-x}$Pd$_x$, 
\cite{CNCJ98,MBL96} and the Kondo-disorder model is in agreement with the 
transport and optical data for UCu$_{5-x}$Pd$_x$ alloys. The Griffiths-phase 
picture describes the thermodynamic properties of a number of NFL materials. 
\cite{dACDD98} 

It should be pointed out that NFL mechanisms based on unconventional Kondo 
effects\cite{Cox87,AfLu91} have to date been treated only for isolated $f$ 
ions. It would be useful to extend such pictures, first to the case of ordered 
$f$-ion-based compounds and then with the inclusion of disorder in analogy with 
the Griffiths-phase model.

Similarities and differences between the Kondo-dis\-order and Griffiths-phase 
theories of NFL behavior and the necessity for their modification at low 
temperatures are discussed in the light of our experimental data. Fits of the 
theories to the temperature dependence of the bulk susceptibility determine the 
parameters of each model, each of which then predicts the temperature 
dependence and size of the NMR linewidth with no further adjustable parameters. 
The measured linewidths are in good agreement with both models. This 
corroborates the conclusion of Graf {\em et al.\/}, \cite{GTHM97} based on 
susceptibility and specific heat measurements, that disorder-driven NFL 
behavior is important in this system. 

\begin{figure}[ht]
\epsfxsize=\columnwidth \epsfbox{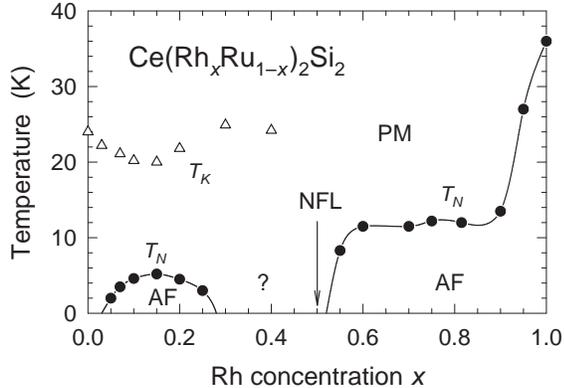} 
\caption{Magnetic phase diagram of Ce(Ru$_{1-x}$Rh$_x$)$_2$Si$_2$. Data from 
Refs.~\protect\onlinecite{LCBE87,SSAM92,SSAM92b}. Paramagnetic (PM) and 
antiferromagnetic (AF) regions are indicated. Circles: AF ordering 
temperatures~$T_N$. Triangles: Kondo temperatures~$T_K$. The curves are guides 
for the eye. The arrow indicates the concentration for which NFL behavior was 
observed by Graf {\em et al.\/} (Ref.~\protect\onlinecite{GTHM97}).}
\label{fig:Crrsfig1}
\end{figure}

The isostructural \mbox{alloy} system~Ce(Ru$_{1-x}$Rh$_x$)$_2$Si$_2$ exhibits a 
variety of complex behavior associated with the Kondo effect and magnetic 
interactions. \cite{HFLL87,LCBE87,SSAM92,SSAM92b,KKFM95,MSA95} The phase 
diagram of this system \cite{LCBE87,SSAM92,SSAM92b} is shown in 
Fig.~\ref{fig:Crrsfig1}. 
The end compound~CeRu$_2$Si$_2$ is a heavy-fermion 
metal which shows no long-range magnetic order down to 20 mK, \cite{HFLL87} 
whereas CeRh$_2$Si$_2$ is an antiferromagnet \cite{LCBE87} with N\'eel 
temperature~$T_N = 36$ K\@. Antiferromagnetism is found for low ($0.1\lesssim x 
\lesssim 0.3$) and high ($0.5 \lesssim x \leq 1$) rhodium doping. Neutron 
diffraction experiments \cite{KKFM95} in the low-doping range show that the 
antiferromagnetism is incommensurate (i.e., the $4f$ electrons are itinerant), 
but becomes commensurate, indicative of local moments, for $x \gtrsim 0.5$. For 
$x \lesssim 0.4$ the specific heat indicates a characteristic energy scale, 
usually associated with the average Kondo temperature~$T_K$ of the material, 
which is higher than $T_N$ in this composition range.

The concentration~$x = 0.5$ is near the critical value for suppression of $T_N$ 
to zero. In spite of its stoichiometric composition CeRhRuSi$_2$ is a 
disordered alloy, as the Rh and Ru atoms occupy the same crystallographic site 
and there is no evidence of superlattice formation. For $x = 0.5$ Graf {\em et 
al.\/} \cite{GTHM97} found the weak divergences characteristic of NFL behavior 
\cite{ITP96} in $\gamma(T)$ and $\chi(T)$ for temperatures between 1 and 30 
K\@. There was no evidence for magnetic ordering, and Graf {\em et al.\/}\ 
concluded that NFL phenomena in CeRhRuSi$_2$ are driven by structural 
disorder. To our knowledge the region~$0.3 \protect\lesssim x \protect\lesssim 
0.5$ has not been examined for NFL behavior. 

Below 1 K $\gamma(T)$ was seen to saturate, suggesting that CeRhRuSi$_2$ 
exhibits a crossover from a region of anomalous magnetic response to a 
Fermi-liquid ground state as the temperature is lowered. It should be noted, 
however, that recent specific heat and ac susceptibility studies of 
UCu$_{5-x}$Pd$_x$ \cite{VMPvL97,SSHP98} suggest that saturation of $\gamma(T)$ 
in this system may be associated with magnetic ordering, possibly of a 
spin-glass nature or in the form of superparamagnetic clusters. More 
information on the behavior of NFL systems at low temperatures is clearly 
needed.

The bulk susceptibility agrees well with the Kondo-disorder model but is 
overestimated by the Griffiths-phase picture at low temperatures. This is not 
surprising, since the possibility of nearly-NFL behavior is built into the 
Kondo-disorder model, whereas the Griffiths-phase theory in its present form 
neglects effects, such as transverse Kondo fluctuations or residual 
interactions between clusters, which could modify or remove the Griffiths 
singularities which cause the NFL behavior.

The experimental data obtained to date do not discriminate clearly between the 
two pictures, since the distribution function $P(\chi)$ which describes the 
inhomogeneous distribution of susceptibilities contains no qualitative feature 
sensitive to the existence of RKKY-coupled clusters. We speculate that 
dynamical properties such as nuclear spin-lattice relaxation rates may be more 
sensitive to low-lying excitations associated with spin-spin couplings, 
particularly at low temperatures, and suggest that further measurements of 
spin relaxation rates be carried out.

Sec.~\ref{sect:exp} of the paper describes our measurements of bulk magnetic 
susceptibility and $^{29}$Si NMR spectra in CeRhRuSi$_2$. The relation between 
inhomogeneity in the susceptibility and the NMR linewidth is reviewed in 
Sec.~\ref{sect:inhomo}. Sec.~\ref{sect:disdriv} treats the single-ion 
Kondo-disorder and Griffiths-phase disorder-driven NFL mechanisms. Analysis of 
the susceptibility and NMR data is discussed in Sec.~\ref{sect:results}, and 
Sec.~\ref{sect:concl} gives our conclusions.

\section{Experiment} \label{sect:exp}

\begin{figure}%
[ht]								
\epsfxsize=\columnwidth \epsfbox{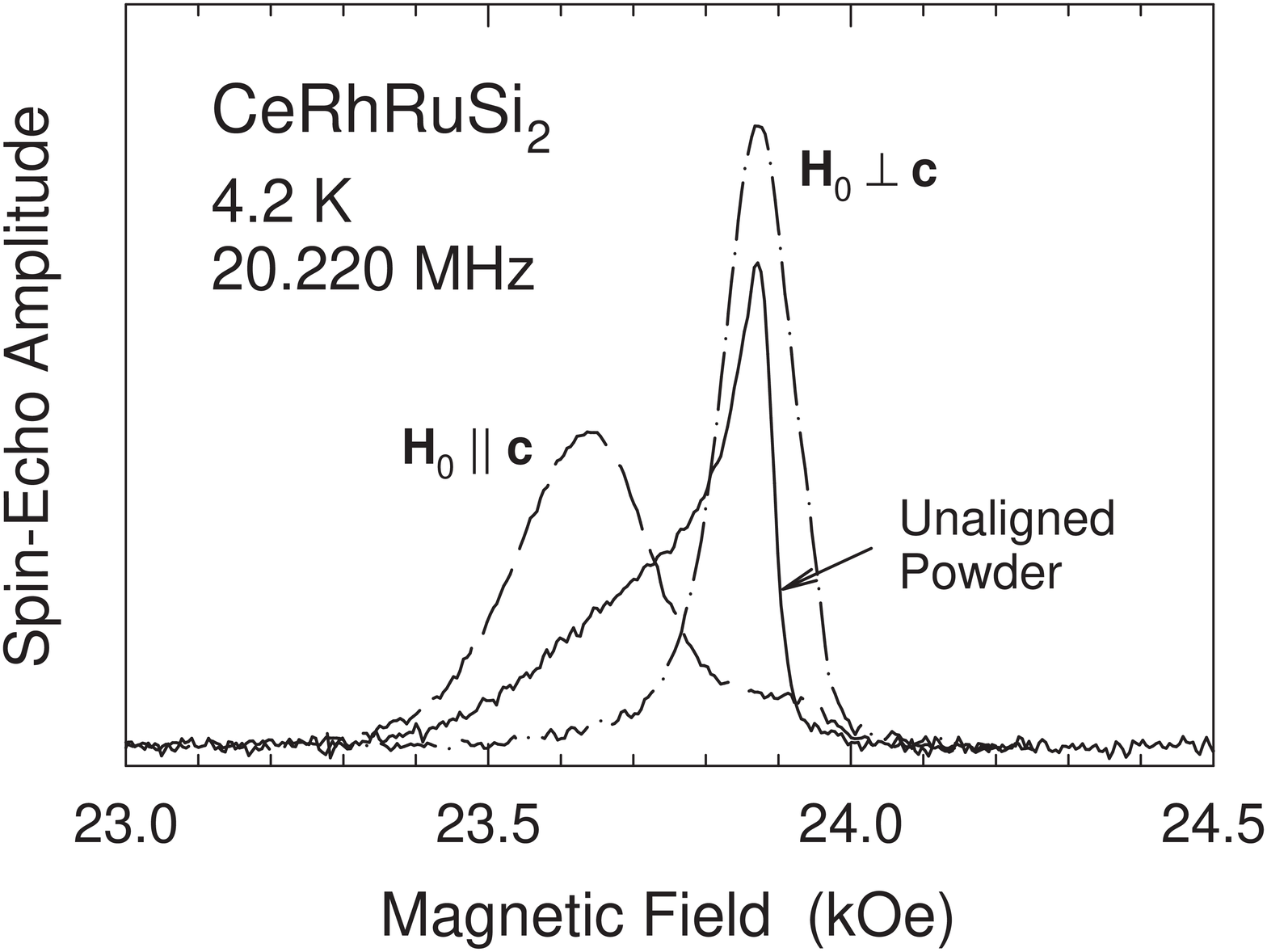}
\caption{Field-swept $^{29}$Si NMR spectra in CeRhRuSi$_2$ for $T=4.2$ K and 
spectrometer frequency 20.220 MHz. Solid curve: unaligned powder sample. Dashed 
curve: field-aligned powder, ${\bf H}_0 \parallel {\bf c}$. Dash-dot curve: 
field-aligned powder, ${\bf H}_0 \perp {\bf c}$.}
\label{fig:Crrsfig2}
\end{figure}

Measurements of bulk susceptibility and $^{29}$Si NMR spectra were carried out 
on unaligned and field-aligned \cite{FCMF87} powder samples of CeRhRuSi$_2$ at 
a frequency of 20.220 MHz and temperatures in the range 4.2--230 K\@. 
Field-swept NMR spectra were obtained using pulsed-NMR spin-echo signals and 
the frequency-shifted-and-summed Fourier-transform processing technique 
described by Clark {\em et al.} \cite{CHLS95} The solid curve in 
Fig.~\ref{fig:Crrsfig2} shows a $^{29}$Si NMR spectrum from an unaligned powder 
sample at 4.2 K\@. 
We attempted to fit this spectrum to an anisotropic powder 
pattern \cite{CBK77} convoluted with a Gaussian broadening function, but found 
a poor fit if the width of the broadening is assumed independent of crystallite 
orientation. The low-field side of the spectrum, which is due to those 
crystallites with $c$ axes parallel to the applied field~${\bf H}_0$ ($H_0 
\approx 24$ kOe), is more strongly broadened than the high-field side. Fits to 
the low-field region of the spectrum yielded a crude estimate of the extra 
broadening, which becomes large at low temperatures as predicted by the 
disorder-driven NFL mechanisms discussed above.

The magnetic susceptibility~$\chi(T)=M(H,T)/H$, where $M(H,T)$ is the bulk 
magnetization of the system, is strongly anisotropic in the 
Ce(Ru$_{1-x}$Rh$_x$)$_2$Si$_2$ series, with the $c$-axis 
susceptibility~$\chi_{c}(T)$ (${\bf H}_0 \parallel {\bf c}$) much larger than 
the $ab$-plane susceptibility~$\chi_{ab}(T)$ (${\bf H}_0 \perp {\bf c}$). 
\cite{LCBE87} This suggests that the extra broadening observed for ${\bf H}_0 
\parallel {\bf c}$ might be due to disorder in the susceptibility similar to 
that found in UCu$_{5-x}$Pd$_x$, and we were motivated to measure the linewidth 
in a field-aligned powder sample. \cite{FCMF87} The powder was mixed with 
epoxy, which was allowed to harden in a magnetic field of 60 kOe. The torque on 
the anisotropic moment aligned the $c$ axis of each single-crystal powder grain 
in the direction of the applied field before the epoxy hardened.

\begin{figure}%
[ht]								
\epsfxsize=\columnwidth \epsfbox{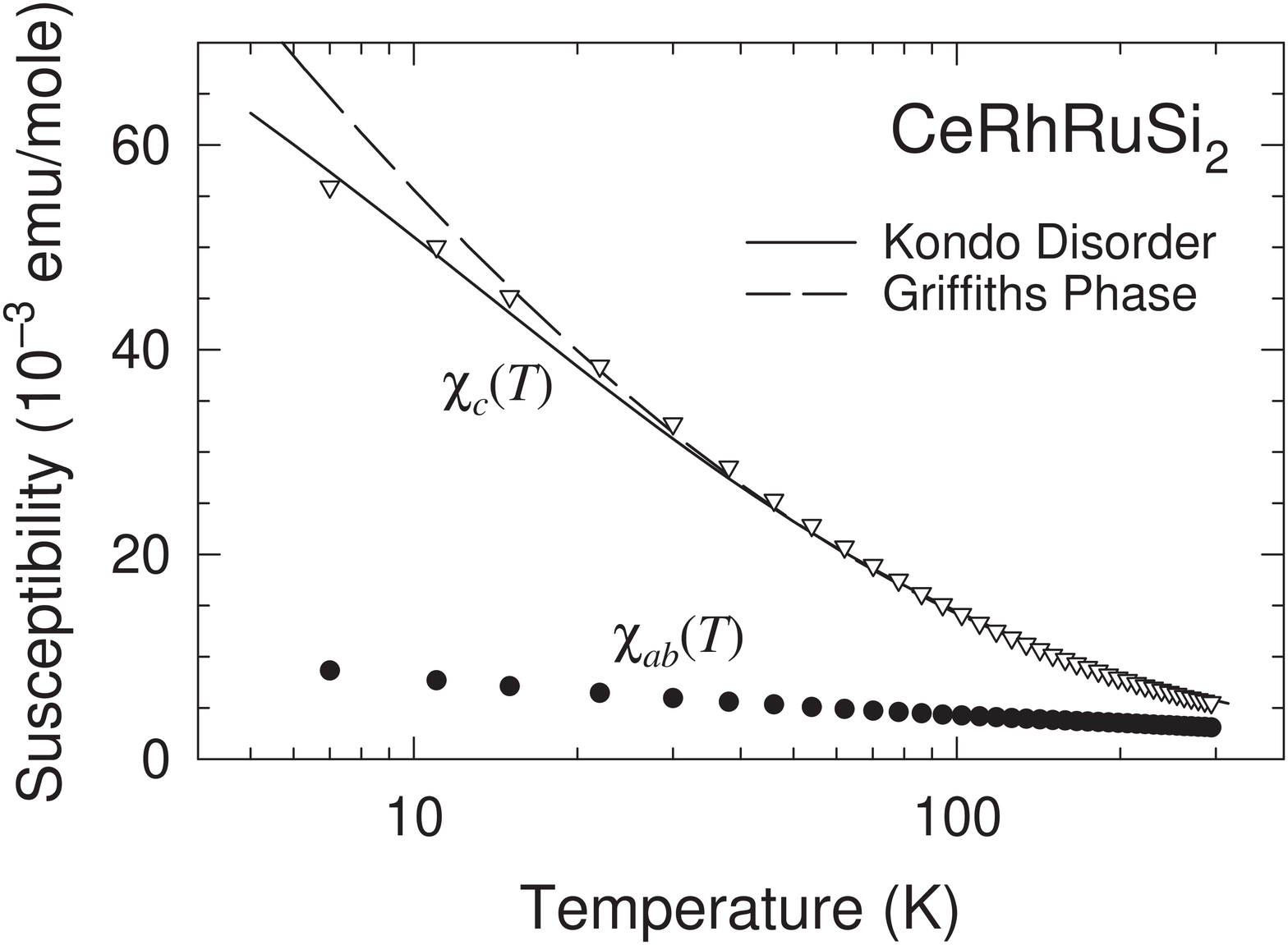}
\caption{Temperature dependence of the anisotropic susceptibility in a 
field-aligned powder sample of CeRhRuSi$_2$. Circles: basal-plane 
susceptibility~$\chi_{ab}(T)$. Triangles: $c$-axis susceptibility~$\chi_c(T)$. 
Solid curve: fit of single-ion Kondo-disorder model 
(Refs.~\protect\onlinecite{BMLA95} and \protect\onlinecite{MBL96}) to 
$\chi_c(T)$. Dashed curve: fit of Griffiths-phase model 
(Ref.~\protect\onlinecite{CNCJ98}) to $\chi_c(T)$.}
\label{fig:Crrsfig3}
\end{figure}

The anisotropic susceptibility measured in this field-aligned powder sample is 
shown in Fig.~\ref{fig:Crrsfig3}. 
These data agree well with measurements on a 
small single crystal of CeRhRuSi$_2$ (not shown). A strong Curie-Weiss-like 
temperature dependence is found for $\chi_c(T)$, whereas $\chi_{ab}(T)$ is 
small and only weakly temperature dependent. \cite{upturn} The curves give fits 
of $\chi_c(T)$ to the Kondo-disorder and Griffiths-phase models as discussed 
below in Sects.~\ref{sect:KD} and \ref{sect:Griff}, respectively.

Figure~\ref{fig:Crrsfig2} also gives the $^{29}$Si NMR spectra measured in the 
field-aligned powder sample for ${\bf H}_0 \parallel {\bf c}$ and ${\bf H}_0 
\perp {\bf c}$. It can be seen that, as expected, the line is wider for ${\bf 
H}_0 \parallel {\bf c}$ than for ${\bf H}_0 \perp {\bf c}$. The small shoulder 
on the high-field side of the ${\bf H}_0 \parallel {\bf c}$ line indicates that 
the alignment of crystallites in this sample is not perfect. The large 
linewidth anisotropy implies, however, that neither a small misalignment of the 
crystallites in the sample nor a slight misalignment of the sample with respect 
to ${\bf H}_0$ affect linewidth measurements appreciably for ${\bf H}_0 
\parallel {\bf c}$. $^{29}$Si NMR spectra from a more completely aligned but 
smaller sample (not shown) confirmed this expectation. The misalignment does, 
however, preclude any attempt to obtain information about the shape of 
$P(\Delta)$ from the shape of the NMR line.

\section{Susceptibility inhomogeneity and NMR linewidth} \label{sect:inhomo} 

Since the NMR frequency shift of a given nucleus is determined by the 
interaction between its magnetic moment and those of the surrounding electrons, 
any spatial variation of the electronic magnetic susceptibility will be 
reflected in the NMR linewidth as a distribution of frequency shifts. A 
quantitative understanding of the susceptibility inhomogeneity requires 
analysis of the relation between it and the NMR linewidth, independent of the 
particular mechanism which causes the inhomogeneity. 

The NMR frequency shift~$K$ measures the time-averaged effective field produced 
by the local moment at the resonating nucleus. In a paramagnet the relative 
shift~$K_i$ of the $i^{\rm th}$ nucleus is related to the local 
susceptibility~$\chi_j$ associated with the $j^{\rm th}$ $f$-ion electronic 
moment by \cite{CBK77}
\begin{equation}
K_i=\sum_ja_{ij}\chi _j \,,
\label{eq:Kandchi}
\end{equation}
where $a_{ij}$ is the hyperfine coupling constant between the $j^{\rm th}$ 
moment and the $i^{\rm th}$ nucleus. It is straightforward to carry out the 
spatial averages and show that 
\[ \overline{K} = a\overline{\chi} \,,\quad a\equiv \sum_j
a_{ij}\,, \]
where a bar designates a spatial average in this and the following. Similarly, 
the rms spread of shifts~%
$\delta K \equiv (\overline{K^2} - \overline{K}\kern.5pt^{\raise1pt\hbox{$\scriptstyle2$}})^{1/2}$ 
is related to the corresponding rms spread of susceptibilities~$\delta\chi$ by 
\[ \delta K = a^{\textstyle*}\delta\chi \,, \]
where $a^{\textstyle*}$ is an effective hyperfine coupling constant, discussed 
in more detail below. As a consequence
\begin{equation}
\delta \chi /\overline{\chi} =
\delta K/(a^{\textstyle*}\overline{\chi}) \,.
\label{eq:Crrsfig7}
\end{equation}

If each nucleus is coupled to more than one moment [cf.\ 
Eq.~(\ref{eq:Kandchi})], any spatial correlation between the moment 
susceptibilities will affect the value of $a^{\textstyle*}$. There are two 
extreme limits in considering this spatial correlation. The term ``long-range 
correlation'' (LRC) will be used to describe the situation where the 
correlation length between local moments is much longer than the local-moment 
near-neighbor spacing. Similarly, ``short-range correlation'' (SRC) describes 
the situation where the variation of susceptibility from site to site is random 
or nearly so, i.e., where the correlation length which describes this variation 
is of the order of or shorter than a lattice constant. Note that this 
correlation is only a phenomenological description of the inhomogeneous 
susceptibility, and is not necessarily related to critical behavior of the 
system. For a given system we do not know {\em a priori\/} which (if either) of 
these limits is applicable, although in the single-ion Kondo-disorder model we 
might expect that random ligand disorder would lead to relatively short-range 
correlation.

It can be shown \cite{MBL96} that the values of $a^{\textstyle*}$ in the LRC 
and SRC limits (denoted by $a_{\rm LRC}^{\textstyle*}$ and 
$a_{\rm SRC}^{\textstyle*}$, respectively) are given by 
\[ a_{\rm LRC}^{\textstyle*} = |a|\,; \quad a_{\rm SRC}^{\textstyle*} = \left( 
\sum_j a_{ij}^2 \right)^{\!\!1/2}. \]
Assuming for simplicity that the hyperfine coupling is predominantly to an 
effective number $n_{\rm eff}$ of $f$-ion near neighbors and is the same effective 
value~$a_{\rm eff}$ for each of these neighbors, it follows that
\[ a_{\rm SRC}^{\textstyle*} = \sqrt{n_{\rm eff}}a_{\rm eff}\quad {\rm and} \quad 
a_{\rm LRC}^{\textstyle*} = n_{\rm eff}a_{\rm eff}\,, \]
so that
\begin{equation}
 a_{\rm SRC}^{\textstyle*} = a_{\rm LRC}^{\textstyle*}/\sqrt{n_{\rm eff}}
\,. 
\label{eq:aSRC}
\end{equation}

In the LRC limit the fractional susceptibility 
inhomogeneity~$\delta\chi/\overline{\chi}$ is given by the relative NMR 
line\-width~$\delta K/\left|\overline{K}\right|$. Since $\delta K = \sigma 
/H_0$, where $\sigma$ is the rms linewidth in magnetic field units, we have 
from Eq.~(\ref{eq:Crrsfig7})
\begin{eqnarray}
\frac{\delta\chi}{\overline{\chi}} & = & \frac{\delta K} {a_{\rm 
LRC}^{\textstyle*} \overline{\chi}} = \frac{\delta K}
{\left|\overline{K}\right|} \nonumber \\
& = & \frac{\sigma}{\left|\overline{K}\right|H_0}\quad \mbox{(LRC limit)} \,. 
\label{eq:LRClim}
\end{eqnarray}
Thus $\sigma /(\left|\overline{K}\right| H_0)$, which can be derived from the 
NMR data, is an estimator for $\delta \chi /\overline{\chi}$ in the LRC limit. 
The corresponding estimator in the SRC limit can be obtained from $\sigma 
/(\left|\overline{K}\right| H_0)$ simply by scaling by the 
factor~$\sqrt{n_{\rm eff}}$ [Eq.~(\ref{eq:aSRC})]:
\begin{eqnarray}
\frac{\delta\chi}{\overline{\chi}} & = & \frac{\delta K}{a_{\rm SRC}^{\textstyle*}\overline{\chi}} = \left(\frac{\delta K}
{\left|\overline{K}\right|}\right) \left( \frac{a_{\rm LRC}^{\textstyle*}}
{a_{\rm SRC}^{\textstyle*}} \right) \nonumber \\ 
& = & \sqrt{n_{\rm eff}}\left(\frac{\delta K}{\left|\overline{K}\right|}\right) 
\quad \mbox{(SRC limit)} \,. 
\label{eq:SRClim}
\end{eqnarray}

The above assumes that the coupling constants~$a_{ij}$ are not disordered, 
i.e., that they have the same values for crystallographically equivalent 
positions of nucleus~$i$ and $f$-ion~$j$. If this is not the case and the 
$a_{ij}$ are also disordered, then it can be shown \cite{MBL96} that
\begin{equation}
\frac{\delta K}{\left|\overline{K}\right|} = \left[ \left( \frac{\delta\chi} {\overline{\chi}} \right)^{\!\!2} + A^2 \right]^{1/2} \,, 
\label{eq:disorderina}
\end{equation}
where $A$ is a term which expresses the effect of the disordered $a_{ij}$. Now 
in existing disorder-driven models \cite{BMLA95,MDK96,CNCJ98} it is found that 
$\delta\chi/\overline{\chi}$ varies considerably with $\overline{\chi}$ (with 
temperature an implicit parameter), tending to a value $\gtrsim 1$ at low 
temperatures and vanishing as $\overline{\chi} \to 0$ (high temperatures). On 
the other hand $A$ is found to be independent of $\overline{\chi}$. Disorder in 
the $a_{ij}$ will therefore result in a nonzero intercept in a plot of $\delta 
K/\left|\overline{K}\right|$ vs.\ $\overline{\chi}$, and its effect can be 
removed by subtracting this intercept in quadrature from the raw $\delta 
K/\left|\overline{K}\right|$ data. It can be shown that this correction is 
valid in both the LRC limit and the SRC limit.

It should be stressed that this ``NMR technology'' is quite independent of the 
specific mechanism which causes the susceptibility inhomogeneity.  

\section{Disorder-driven NFL mechanisms} \label{sect:disdriv}

In considering systems where NFL behavior is driven by disorder it is 
convenient to study the spatially-distributed local susceptibility~$\chi_j$. 
Simple linear response theory shows that the zero-temperature local 
susceptibility can be associated with a characteristic local energy 
scale~$\Delta_j$ by 
\begin{equation}
\chi_j \propto \frac{1}{\Delta_j} \,,
\label{eq:chidelta}
\end{equation}
where $\Delta_j$ is essentially the excitation energy from the ground state to 
the first excited state. At finite temperatures~$T$ the local 
susceptibility~$\chi(\Delta,T)$ depends strongly on the microscopic details 
which couple the magnetic degrees of freedom. 

We can therefore speak of distributions of susceptibilities or energy scales, 
characterized by distribution functions~$P(\Delta)$ and $P(\chi)$, 
respectively; thus $P(\chi,T) = P(\Delta)/|\partial\chi(\Delta,T)/
\partial\Delta|$. Once $P(\Delta)$ [or $P(\chi,T)$] is known we can obtain 
spatial averages of physical quantities such as the $n^{\rm th}$ 
moment~$\overline{\chi^n}(T)$ of the local susceptibility distribution, which 
is given by
\begin{mathletters}
\label{eq:moments}
\begin{equation}
\overline{\chi^n}(T) = \int_0^\infty \chi^n(\Delta,T) P(\Delta)\, d\Delta 
\end{equation}
\vspace*{-10pt} \begin{equation}
\phantom{\chi^n} = \int_0^\infty \chi^n P(\chi,T)\, d\chi \,. 
\end{equation}
\end{mathletters}
Knowledge of the first and second moments of $P(\chi,T)$ is sufficient for 
interpretation of the bulk susceptibility and NMR linewidth data.

\subsection{Single-ion Kondo disorder} \label{sect:KD}

We take the Ce-ion spins to be coupled to the conduction-electron spins by a 
$s$-$f$ exchange coupling described by a coupling constant~$g = N(E_F) J$, 
where $N(E_F)$ is the density of conduction-electron states at the Fermi 
surface and $J$ is the Ce-ion--conduction-electron exchange interaction. If the 
system is disordered on the ligand sites, as in CeRhRuSi$_2$, $g$ will be 
randomly distributed according to a distribution function~$P(g)$. In the 
simplest picture of the Kondo effect, the Kondo temperature $T_K$, which 
characterizes the energy scale of the single-ion Kondo effect, is given by $T_K 
= E_F\, e^{-1/g}$, where $E_F$ is the Fermi energy. Thus a narrow distribution 
of $g$ can lead to a wide distribution of $T_K$ when $g$ is small. In this 
picture we immediately identify $\Delta$ with $T_K$.

If the distribution function~$P(\Delta)=P(T_K)$ is broad enough so that 
$P(T_K{\to}0)$ does not vanish, then at any nonzero temperature~$T$ those $f$ 
ions for which $T_K < T$ are not compensated (i.e., are not described by 
Fermi-liquid theory) and give rise to the NFL behavior. In view of 
Eq.~(\ref{eq:chidelta}) one sees that regions of the system where $T_K$ is very 
small (sites with very large low-temperature susceptibility) dominate the 
thermal and transport properties. Miranda {\em et al.\/} \cite{MDK96} have 
treated this picture in detail, and have shown that it predicts the observed 
low-temperature behavior of the Sommerfeld coefficient~$\gamma(T)$, 
susceptibility~$\chi(T)$, and resistivity~$\Delta\rho(T) = \rho(T) - \rho(0)$ 
($\gamma \propto \chi \propto -\ln T$ and $\Delta\rho \propto T$, respectively) 
provided that $P(T_K{\to}0)$ is finite. 

The resulting distribution function~$P(T_K)$ is given by
\begin{equation}
P(T_K)=P(g)\left| \frac{dg}{dT_K} \right| = \frac{g^2P(g)}{T_K}
\label{eq:P(TK)}
\end{equation}
with $g = 1/\ln(E_F/T_K)$.
As a convenient parameterization of the Kondo physics we take the 
susceptibility to have the Curie-Weiss form
\begin{equation}
\chi(T,T_K) = {\cal C}/(T+\alpha T_K) \,,
\label{eq:chikon}
\end{equation}
where ${\cal C}$ is the Curie constant. The value of $\alpha$ was estimated by 
comparing this Curie-Weiss law to the exact Bethe-ansatz solution; \cite{RLA82} 
the two functional forms differ by ${\lesssim}10$\% for $\alpha \approx 2.5$. 
Assuming a Gaussian distribution for $P(g)$, the mean~$\overline{g}$ and rms 
width~$\delta g$ of the distribution can be found by fitting 
Eq.~(\ref{eq:moments}) with $n = 1$ to the measured bulk (i.e., spatially 
averaged) susceptibility. \cite{BMLA95,GTHM97}

\subsection{Griffiths-phase model} \label{sect:Griff}

In the Griffiths-phase model of NFL behavior \cite{CNCJ98} the low-energy 
physics is dominated by rare and large clusters which can tunnel over 
classically forbidden regions. These correlated regions are generated by 
above-average values of the RKKY interaction. The tunneling is produced by the 
spin-flip processes present in the Kondo effect. \cite{CNCJ98} In this scenario 
the Griffiths singularities appear close to a QCP below percolation threshold 
and are therefore intrinsically related to QCP physics. It is intuitively clear 
that the clusters can be effectively described in terms of two level systems, 
with tunneling energy~$E$ which is distributed over the sample due to the 
structural disorder. Obviously we have $\Delta = E$ in this picture. 

The distribution of $E$ is obtained by mapping the problem onto the Ising model 
in a transverse field; this procedure is valid in the limit of 
large magnetic anisotropy as appears to be the case in CeRuRhSi$_2$ (cf.\ 
Fig.~\ref{fig:Crrsfig3}). Then it can be shown \cite{AHCN98} that 
\begin{equation}
P(E) = \left\{ \begin{array}{ll}
\displaystyle \frac{\lambda}{\epsilon_0} \left( \frac{E}{\epsilon_0} 
\right)^{\!\!-1+\lambda}, & 0 < E < \epsilon_0 \,, \\
\noalign{\vskip5pt}
0 \,, & E > \epsilon_0 \,,
\end{array} \right.
\label{eq:grifprob}
\end{equation}
where $\lambda$ is an exponent that determines the behavior of the response 
functions ($0 \le \lambda \le 1$), and $\epsilon_0$ is a high energy cut-off 
which must be determined for each specific system. 

As discussed above the local zero-temperature susceptibility in this picture is 
$\chi(0,E) = {\cal C}/E$; large clusters with small energy scales have large 
susceptibilities. At high temperatures one expects the clusters to be 
disordered and behave paramagnetically, resulting in a Curie behavior 
$\chi(E,T) = {\cal C}/T$ for $T \gg E$. We therefore assume a Curie-Weiss 
interpolation formula
\begin{equation}
\chi(E,T) = \frac{\cal C}{T + E} 
\label{eq:chigrif}
\end{equation}
as for the Kondo-disorder model. But in the present case this approximation is 
intended to incorporate all the interaction effects which determine the 
susceptibility of a multi-ion cluster, not just single-ion Kondo physics.

Using Eqs.~(\ref{eq:moments}), (\ref{eq:grifprob}), and (\ref{eq:chigrif}) it 
is straightforward to show\cite{AHCN98} that for $T \ll \epsilon_0$
\begin{equation}
\overline{\chi}(T) = \frac{\pi\lambda {\cal C}}{\epsilon_0} \left( 
\frac{\epsilon_0}{T} \right)^{\!\!1-\lambda} \\ 
\label{eq:griffchi}
\end{equation}
and
\begin{equation}
\frac{\delta\chi(T)}{\overline{\chi}(T)} = \left[ \frac{1-\lambda}{2\pi\lambda} 
\right]^{1/2} \left( \frac{\epsilon_0}{T} \right)^{\!\lambda/2} \,.
\label{eq:griffCrrsfig7}
\end{equation}
The critical behavior is determined by the single nonuniversal 
temperature-independent exponent $\lambda$. For $\lambda <1$ the susceptibility 
diverges algebraically as $T \to 0$ ($H = 0$), and NFL behavior is obtained. 
The divergence increases (i.e., the NFL behavior becomes stronger) with 
decreasing $\lambda$.  The case~$\lambda = 1$ is marginal, and leads to 
logarithmic singularities as in the Kondo-disorder approach. \cite{MDK96}\,

\subsection{Kondo disorder versus Griffiths singularities}

Both the Kondo-disorder and Griffiths-singularities pictures deal with a 
similar aspect of disorder, viz., the physics of rare events with large 
susceptibilities. It is clear, however, that the microscopic aspects of the two 
models are very different. The Kondo-disorder model uses non-interacting 
single-ion physics, and no aspect of the RKKY interaction is present. The 
Griffiths-phase approach, on the other hand, tries to take both RKKY and Kondo 
phenomena into account on an equal footing, and has a strong connection with 
QCP physics.

From the point of view of local properties as measured in NMR spectra these 
approaches give similar results. The spatial properties of the two approaches 
are very different, however, since the formation of clusters in the Griffiths 
phase requires spatially extended structure. In this case one could look for 
the existence of clusters via superparamagnetic response, which is well 
understood in the context of spin glasses, \cite{Mydo93} or for a momentum 
dependence of the inelastic neutron scattering. In addition, one would expect 
cluster formation to slow down the spin fluctuations relative to the free-ion 
fluctuation rate, which is essentially $T_K$. Experiments that are sensitive to 
the fluctuation rate may therefore be able to distinguish between the two 
theories.

\section{Results and Discussion} \label{sect:results}

\subsection{Bulk magnetic susceptibility}

A fit of the Kondo-disorder model result for $\overline{\chi}(T)$ 
[Eq.~(\ref{eq:moments}) with $n = 1$ and $\Delta = T_K$, and 
Eqs.~(\ref{eq:P(TK)}) and (\ref{eq:chikon}) for $P(T_K)$ and $\chi(T,T_K)$] to 
the experimental $c$-axis susceptibility~$\chi_c(T)$ is shown as the solid 
curve in Fig.~\ref{fig:Crrsfig3}. We obtain the same coupling constant 
distribution width~$\delta g = 0.021$ as Graf {\em et al.\/}, \cite{GTHM97} and 
a somewhat smaller mean~$\overline{g} = 0.160$ compared to 0.175 from 
Ref.~\onlinecite{GTHM97}. \cite{Graf&us} The coupling constants are less widely 
distributed than in the NFL system~UCu$_{5-x}$Pd$_x$, $x = 1.0$ and 1.5, 
\cite{MBL96} consistent with weaker ``nearly NFL'' behavior in CeRhRuSi$_2$.

Figure~\ref{fig:Crrsfig3} also gives the Griffiths-phase model prediction for 
$\chi_c(T)$ (dashed curve), obtained by fitting $\overline{\chi}(T)$ from 
Eq.~(\ref{eq:moments}) ($n = 1$), with $\Delta = E$ and using 
Eqs.~(\ref{eq:grifprob}) and (\ref{eq:chigrif}), to the bulk $c$-axis 
susceptibility. The best fit is given by the dashed curve in 
Fig.~\ref{fig:Crrsfig3}. It can be seen that at low temperatures the 
Griffiths-phase fit curve overestimates the experimental data slightly. This is 
to be expected, since in the simple Griffiths-phase model there is no 
possibility of a return to Fermi-liquid behavior at low temperatures: the 
system is a true NFL as long as $\lambda < 1$. But the susceptibility data 
begin to exhibit the saturation expected from the conclusions of Graf {\em et 
al.\/},\cite{GTHM97} and therefore are not well described by an algebraically 
divergent temperature dependence. There is, however, a region of intermediate 
temperatures in which both Kondo-disorder and Griffiths-singularity models 
agree very well with experiment. From the Griffiths-phase fit in this region we 
obtain $\epsilon_0 = 170 \pm 10$ K and $\lambda = 0.88 \pm 0.02$. The latter 
value is considerably larger (i.e., the NFL behavior is weaker) than found in 
UCu$_{5-x}$Pd$_x$ \cite{CNCJ98} as in the Kondo disorder model. 

\begin{figure}%
[ht]								
\epsfxsize=\columnwidth \epsfbox{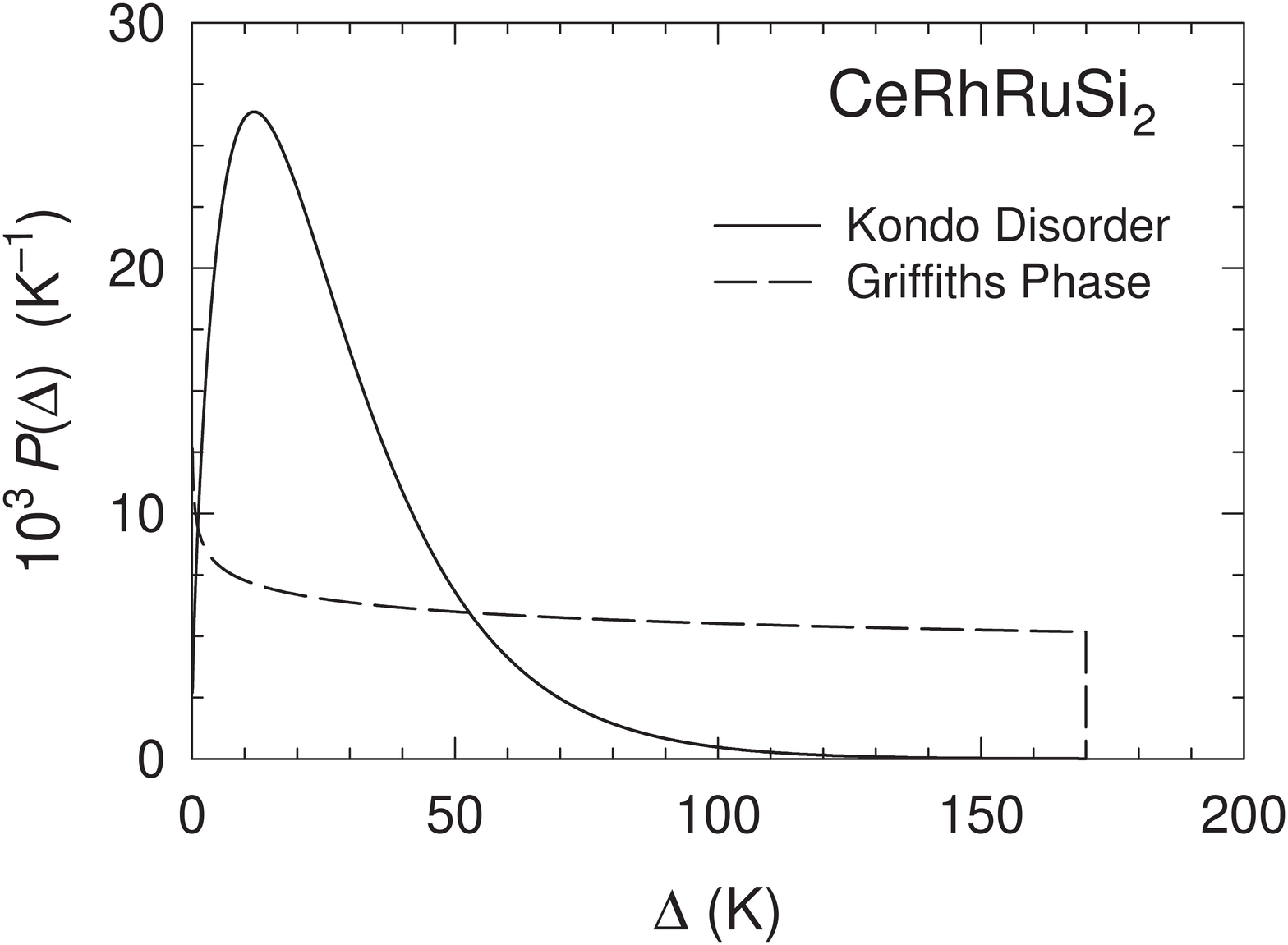}
\caption{Distribution functions $P(\Delta)$ of characteristic energies $\Delta$ 
in CeRhRuSi$_2$, obtained from fitting the Kon\-do-disorder ($\Delta = T_K$) and 
Griffiths-phase ($\Delta = E$) mod\-els to the bulk $c$-axis susceptibility 
(Fig.~\protect\ref{fig:Crrsfig3}). Solid curve: Kondo-disorder distribution 
function~$P(T_K)$ [Eq.~(\protect\ref{eq:P(TK)})]. Dashed curve: Griffiths-phase 
distribution function~$P(E)$ [Eq.~(\protect\ref{eq:grifprob})].}
\label{fig:Crrsfig4}
\end{figure}

Figure~\ref{fig:Crrsfig4} shows the distribution functions~$P(\Delta)$ which 
result from the Kondo-disorder ($\Delta = T_K$) and Griffiths-phase ($\Delta = 
E$) model fits. 
It can be seen that the two functions are very different. The 
Kondo-disorder distribution function~$P(T_K)$ exhibits a maximum near 12 K and 
is small below $\sim$1 K and above $\sim$100 K, whereas the Griffiths-phase 
distribution function~$P(E)$ is broader and diverges weakly as $E \to 0$. These 
differences are much more marked in $P(\Delta)$ than in the corresponding fits 
to the susceptibility (Fig.~\ref{fig:Crrsfig3}); spatially-averaged experimental 
quantities are insensitive to the exact form of $P(\Delta)$. It is clear that 
the Griffiths-phase model could fit the data better if $P(E) \to 0$ as $E \to 
0$, which in an extended Griffiths-phase picture would occur if there were an 
upper cutoff on the cluster susceptibility.

Thus CeRhRuSi$_2$ does not exhibit ``true'' NFL behavior. In the Kondo-disorder 
model, which shows this most explicitly, the best fit indicates that all spins 
are in a Kondo-compensated Fermi-liquid state at low enough temperatures. This 
is consistent with the results of Graf {\em et al.\/} \cite{GTHM97} that 
$\gamma(T) \to {\rm const.}$\ and $\Delta\rho(T) \propto T^2$ as $T \to 0$. We 
note again, however, that as mentioned in Sec.~\ref{sect:intro} the saturation 
of $\gamma(T)$ does not necessarily indicate Fermi-liquid behavior at low 
temperatures; other physics, such as magnetic freezing, \cite{VMPvL97,SSHP98} 
may be involved.

\subsection{$^{29}$Si NMR linewidths}

\begin{figure}%
[ht]								
\epsfxsize=\columnwidth \epsfbox{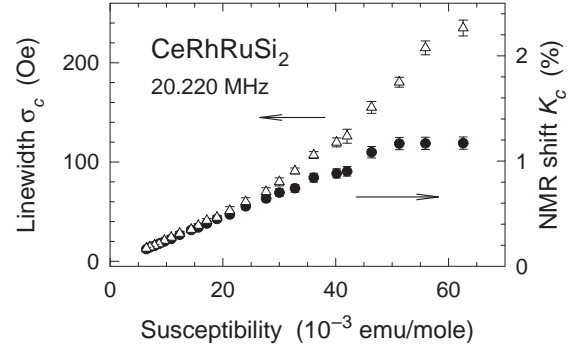}
\caption{$^{29}$Si $c$-axis NMR shift $K_c$ (circles) and linewidth~$\sigma_c$ 
(triangles) vs.\ $c$-axis susceptibility~$\chi_c$ in CeRhRuSi$_2$, with 
temperature an implicit parameter. Spectrometer frequency 20.220 MHz (applied 
field $\sim 23.9$ kOe). The linewidth~$\sigma_c$ varies more rapidly with 
$\chi_c$ than $K_c$, as expected from disorder-driven theories of NFL behavior 
(Refs.~\protect\onlinecite{MDK96,CNCJ98,BMLA95}).}
\label{fig:Crrsfig5}
\end{figure}

The $^{29}$Si $c$-axis NMR shift~$K_c$ and linewidth~$\sigma_c$ are plotted 
against the $c$-axis bulk susceptibility $\chi_c$ in Fig.~\ref{fig:Crrsfig5}, with 
temperature an implicit parameter. 
(The $ab$-plane parameters~$K_{ab}$ and 
$\sigma_{ab}$, not shown, are small and only weakly temperature dependent.) It 
can be seen that $\sigma_c$ varies more rapidly with $\chi_c$ than $K_c$, as 
expected qualitatively from disorder-driven theories of NFL behavior. 
\cite{MDK96,CNCJ98,BMLA95} Although the expected linear relation between $K_c$ 
and $\chi_c$ is observed at high temperatures (small $\chi_c$), $K_c(\chi_c)$ 
tends to a constant for large $\chi_c$. This saturation is not well understood, 
but may be due to a small amount of second phase; this could have a strong 
Curie-Weiss-like susceptibility but little effect on the NMR shift since the 
number of nuclei in the second phase would be small.\cite{MacL85} The observed 
nonlinearity is not more than $\sim$20\% and does not affect our conclusions 
significantly.

\begin{figure}%
[ht]								
\epsfxsize=\columnwidth \epsfbox{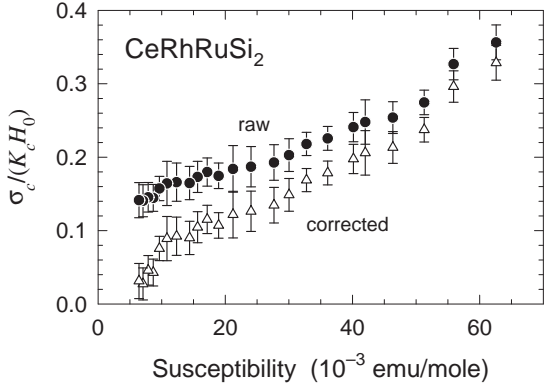}
\caption{Plot of relative spread~$\sigma_c/(K_cH_0)$ of $c$-axis $^{29}$Si NMR 
shifts in LRC limit versus $c$-axis susceptibility $\chi _c$ in CeRhRuSi$_2$. 
See text for symbol definitions. Circles: raw data before correction for 
disorder in the hyperfine coupling constant. Triangles: data corrected for 
coupling constant disorder (see text).}
\label{fig:Crrsfig6}
\end{figure}

Figure~\ref{fig:Crrsfig6} plots $\sigma_c/(K_cH_0) = \delta K_c/K_c = \delta K_c/
(a_{\rm LRC}^{\textstyle*}\chi_c)$ [Eq.~(\ref{eq:LRClim})] versus $\chi_c$. 
As discussed in Sec.~\ref{sect:inhomo} $\sigma_c/(K_cH_0)$ is an estimator for 
$\delta\chi/\overline{\chi}$ in the LRC limit. The data extrapolate to a 
non-zero value as $\chi_c \to 0$, which indicates that the coupling 
constant~$a_{ij}$ is also disordered. We therefore subtracted the extrapolated 
intercept from the raw values in quadrature [cf.\ Eq.~(\ref{eq:disorderina})] 
to obtain corrected data in the LRC limit, also shown in Fig.~\ref{fig:Crrsfig6} 
(triangles). These corrected data represent $\delta K_c/(a_{\rm LRC}^
{\textstyle*}\chi_c)$ with $\delta K_c$ due only to susceptibility 
inhomogeneity. The corresponding (corrected) values of $\delta K_c/(a_{\rm 
SRC}^{\textstyle*}\chi_c)$ were obtained from Eq.~(\ref{eq:SRClim}) of 
Sec.~\ref{sect:inhomo}, with $n_{\rm eff}$ chosen as described below.

\begin{figure}%
[ht]								
\epsfxsize=\columnwidth \epsfbox{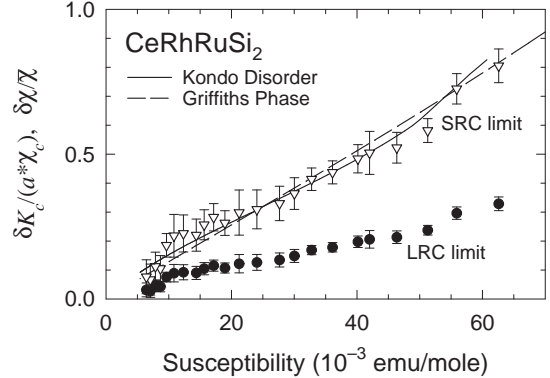}
\caption{Experimental and theoretical relations between rms susceptibility 
inhomogeneity and  bulk susceptibility in CeRhRuSi$_2$. Data points: $c$-axis 
NMR estimator~$\delta K_c/(a^{\textstyle*}\chi_c)$ of relative rms 
susceptibility spread~$\delta\chi/\overline{\chi}$. Circles: LRC limit (see 
text). Triangles: SRC limit. Solid curve: $\delta\chi/\overline{\chi}$ in the 
single-ion Kondo-disorder model. Dashed curve: $\delta\chi/\overline{\chi}$ in 
the Griffiths-phase model. Agreement is found between predictions of 
disorder-driven NFL theories and data in the SRC limit for effective 
nearest-neighbor number~$n_{\rm eff} = 6$.}
\label{fig:Crrsfig7}
\end{figure}

For both the Kondo-disorder and Griffiths-phase models $\delta\chi/
\overline{\chi}$ was calculated from Eq.~(\ref{eq:moments}) ($n = 2$) and 
$\overline{\chi}(T)$ with no further adjustable parameters, since $P(\Delta)$ 
had been previously determined by the fits to the bulk susceptibility. 
Figure~\ref{fig:Crrsfig7} compares $\delta K_c/(a^{\textstyle*}\chi_c)$ in both 
limits with the theoretical behavior of $\delta\chi/\overline{\chi}$ from these 
theories, again with temperature an implicit parameter. 
It can be seen that the 
theoretical predictions are similar and that they both overestimate $\delta 
K_c/(a_{\rm LRC}^{\textstyle*} \chi_c)$ considerably, but that the agreement 
with $\delta K_c/(a_{\rm SRC}^{\textstyle*} \chi_c)$ is excellent when $n_{\rm 
\rm eff}$ in Eq.~(\ref{eq:SRClim}) is taken to be 6. 

\begin{figure}%
[ht]								
\epsfxsize=\columnwidth \epsfbox{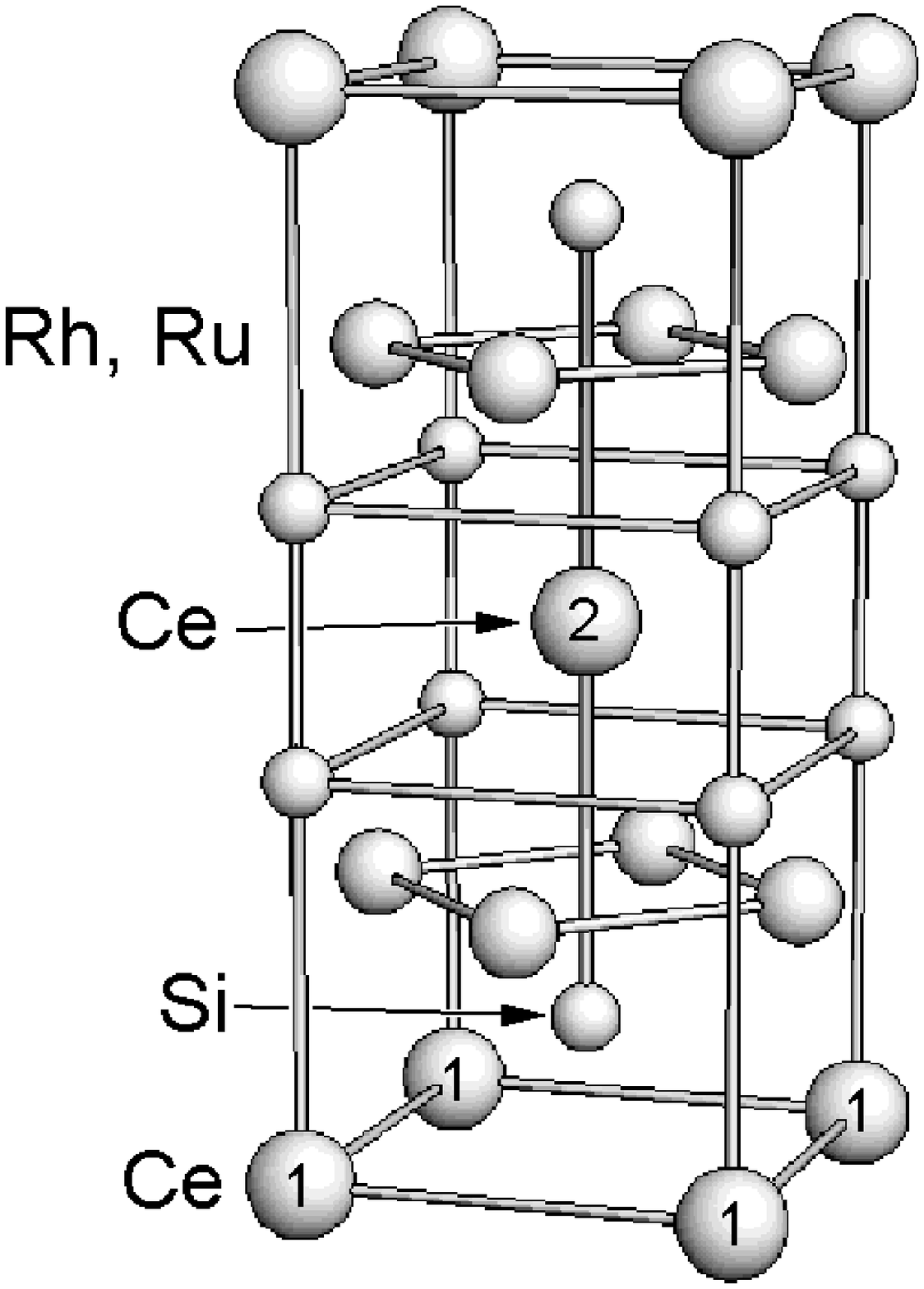}
\caption{Crystal structure of Ce(Rh$_{1-x}$Ru$_x$)Si$_2$. Each Si site is 
coordinated by four Ce nearest neighbors (no.~1) in the adjacent basal plane 
and one Ce next-nearest neighbor (no.~2) in the opposite basal plane.}
\label{fig:Crrsfig8}
\end{figure}

That this value is sensible can be concluded from examination of the 
Al$_4$Ba-type crystal structure of CeRhRuSi$_2$, shown in 
Fig.~\ref{fig:Crrsfig8}, where it can be seen that each Si site is coordinated 
by four Ce nearest neighbors and one Ce next-nearest neighbor. 
Thus $n_{\rm eff}$ is approximately 
the coordination number for the first two near-neighbor shells and is therefore 
reasonable, given the approximation of an effective number of equally-coupled 
neighbors.

For CeRhRuSi$_2$ we do not have the independent verification of the SRC limit 
that was available from comparison of NMR and muon spin rotation ($\mu$SR) 
spectra in the case of UCu$_{5-x}$Pd$_x$. \cite{BMAF96} (For a review of the 
$\mu$SR technique see, for example, Ref.~\onlinecite{Sche85}.) The latter 
alloys have a cubic crystal structure and the positive-muon ($\mu^+$) 
interstitial stopping sites possess octahedral and tetrahedral point 
symmetries, which are sufficiently high to render the $\mu^+$ frequency shift 
isotropic. Then the $\mu^+$ linewidth reflects the susceptibility inhomogeneity 
rather than anisotropic powder-pattern broadening. Preliminary $\mu$SR 
measurements in an unaligned powder sample of CeRhRuSi$_2$ \cite{MHLT98} show 
that in this alloy the anisotropic contribution dominates the powder-pattern 
linewidth, much as in the unaligned-powder spectrum of Fig.~\ref{fig:Crrsfig2}, 
and the disorder-induced broadening cannot be determined accurately. 

Unfortunately field-aligned powder samples cannot be used in $\mu$SR 
experiments. The packing fraction of the powder must be small ($\lesssim 20$\%) 
in order to allow free rotation of the powder grains during alignment, and then 
only a correspondingly small fraction of the muons stop in the sample; the rest 
stop in the epoxy and give a spurious signal. Thus we cannot confirm the SRC 
limit by comparing results between NMR and $\mu$SR. We also note that no other 
nucleus in CeRhRuSi$_2$ is favorable for NMR; stable Ce isotopes possess no 
nuclear magnetic moment, and Ru and Rh isotopes have very small gyromagnetic 
ratios. Nevertheless, the SRC-limit estimate of $\delta\chi_c/\chi_c$ is in 
excellent self-consistent agreement with the disorder-driven models.

\section{Conclusions} \label{sect:concl}

The picture that emerges from our $^{29}$Si NMR stu\-dy of CeRhRuSi$_2$ 
exhibits similarities and differences when compared to the preceding NMR 
investigation of UCu$_{5-x}$Pd$_x$, $x = 1.0$ and 1.5. \cite{BMLA95,MBL96} The 
most important similarity is the fact that in both cases the NMR data are in 
excellent agreement with predictions of disorder-driven theories of NFL 
behavior. Our results therefore confirm the conclusions of Graf {\em et al.\/} 
\cite{GTHM97} that such a mechanism drives NFL properties in CeRhRuSi$_2$. The 
most important differences between the two systems are that in CeRhRuSi$_2$ 
(1)~within the single-ion Kondo-disorder model the disorder is not enough to 
prevent a return to Fermi-liquid behavior at temperatures $\lesssim 1$ K, and 
(2)~the determination of the appropriate correlation length limit (LRC or SRC) 
has not been made independently of comparison with theory. The agreement 
between theory and experiment assuming the SRC limit (Fig.~\ref{fig:Crrsfig7}) is 
satisfactory.

From the experimental point of view, the relatively small differences between 
the predictions of the single-ion Kondo-disorder picture and the 
Griffiths-phase theory show how difficult it is to discriminate between these 
two mechanisms for disorder-driven NFL behavior in the NMR linewidths. We 
speculate, however, that the dynamics of the spins will be quite different in 
the two cases, particularly at low temperatures. 

The single-ion Kondo disorder model predicts inhomogeneous relaxation due to 
the distributed $T_K$. This mechanism yields a spatially averaged spin-lattice 
relaxation rate~$\overline{T_1^{-1}(T)} = \int dT_K\,P(T_K) T_1^{-1}(T,T_K)$. 
It is convenient to approximate $T_1^{-1}(T,T_K)$ by
\[ T_1^{-1}(T,T_K) \propto \left\{ \begin{array}{ll}
1/T_K\,, & T > T_K \,, \\
\noalign{\vskip4pt}
T/T_K^2\,, & T < T_K \,, \\
\end{array} \right.
\]
which captures the crossover to Fermi-liquid (Korringa) behavior for $T < T_K$. 
For a model rectangular $P(T_K)$ given by
\[ P(T_K) = \left\{ \begin{array}{cl}
\displaystyle \frac{1}{T_M - T_m} \,, & T_m < T_K < T_M \,, \\
\noalign{\vskip4pt}
0 \,, & \mbox{otherwise,} \\
\end{array} \right.
\]
where $T_m$ and $T_M$ are minimum and maximum values of $T_K$, respectively, it 
is straightforward to show that $\overline{T_1^{-1}}$ varies linearly with $T$ 
for $T < T_m$ and goes smoothly to a constant for $T > T_M$. Such behavior 
would be generally expected to characterize $\overline{T_1^{-1}}$ as long as 
$P(T_K{\to}0) = 0$. Thus in this scenario $\overline{T_1^{-1}}$ depends on 
temperature but is independent of resonance frequency~$\omega$. 

In contrast, it can be easily shown from the expression for the dissipative 
dynamic susceptibility~$\chi''(\omega)$ in the Griffiths-phase theory 
\cite{CNCJ98} 
\[ \chi''(\omega) \propto \omega^{-1+\lambda} \tanh(\hbar\omega/k_BT) \]
that, assuming the validity of this picture for the very low nuclear (muon) 
frequencies~($\hbar\omega \ll k_BT$)
\[ \overline{T_1^{-1}} \propto \omega^{-1+\lambda} \,, \]
independent of temperature. The frequency is given by $\omega = \gamma H_0$, 
where $\gamma$ is the nuclear (muon) gyromagnetic moment. Thus the dependence 
of $\overline{T_1^{-1}}$ on temperature and $H_0$ differs consierably between 
the two theories.

It should be noted that the NMR linewidths obtained from experimental data are 
not necessarily true sample rms averages, if the lines have extended shoulders 
which are lost in the noise and not taken into account. The fact that good fits 
are obtained with Gaussian lines seems to make this unlikely, but it is not 
hard to see that at low temperatures $P(\chi)$ should be broad and asymmetric 
in both the Kondo-disorder and Griffiths-phase models. If the experimental 
linewidth characterizes only ``typical'' environments it will underestimate the 
true sample average. This would render our quantitative results somewhat 
uncertain, but would not invalidate the conclusion that disorder is an 
important element in the NFL behavior of CeRhRuSi$_2$ above 1 K. 

Finally, we discuss the relation of the disorder-driven theories to the 
observed crossover to a new regime (Fermi-liquid behavior, cluster formation, 
magnetic freezing, etc.)\ in CeRuRhSi$_2$ below 1 K\@. The crossover is 
described empirically by the Kondo-disorder model, which by itself gives no 
clue as to why there should (or should not) be a suppression of low Kondo 
temperatures. Recently, however, Miranda and Dobrosavljvi\'c\cite{MiDo99} have 
reported a microscopic calculation of the form of $P(T_K)$ for various levels 
of disorder. They find that the distribution is singular only for sufficiently 
strong disorder, whereas for slightly weaker disorder $P(T_K) \to 0$ at small 
$T_K$, onsistent with a return to Fermi-liquid behavior at the lowest 
temperatures. This feature is in at least qualitative agreement with our 
results.

To explain the crossover the Griffiths-phase scenario would have to be extended 
beyond its simplest form to include a mechanism which reduces the response of 
the largest clusters. The mechanism behind such a reduction could be the 
breakdown of the assumption of strong single-ion anisotropy made in the 
Griffiths-phase theory; \cite{AHCN98} transverse fluctuations of the Ce ions 
might constitute a damping mechanism which rounds off the Griffiths 
singularities. Alternatively, superparamagnetic spin freezing of the clusters 
could occur at very low temperatures. \cite{VMPvL97,SSHP98} 

It is possible that similar crossovers occur in other NFL materials, perhaps at 
temperatures which have not yet been explored. In any event, our experimental 
findings indicate that further development of current theories of 
disorder-driven NFL behavior is required to understand NFL phenomena at low 
temperatures.

\medskip One of us (D.E.M.) is grateful for discussions with C.~H. Booth, R.~H. 
Heffner, M.~F. Hundley, and R. Modler. This research was supported by the 
U.S.~NSF, Grant no.~DMR-9418991 (U.C. Riverside), by the Research Corporation 
(Whittier College), and by the U.C.~Riverside Academic Senate Committee on 
Research, and was performed in part under the auspices of the U.S.~DOE (Los 
Alamos). One of us (A.H.C.N.) acknowledges support from the Alfred P. Sloan 
Foundation.


\begin{thebibliography}{10}
\bibitem{upturn}
The small upturn of $\chi_{ab}(T)$ at low temperatures could result from slight
  misalignment of the powder grains.
\end{thebibliography}
\end{document}